\documentclass[12pt]{article}
\begin{document}

\setlength{\baselineskip}{24pt}

\title{
{\Large \bf The Fermion Determinant, its Modulus and Phase
}}

\author{
Hisashi {\sc Kikuchi}
\\
{\normalsize\sl Ohu University}\\ 
{\normalsize\sl Koriyama 963, Japan}\\
{\normalsize\tt (kikuchi@yukawa.kyoto-u.ac.jp)}
}

\date{}

\maketitle

\begin{abstract}
We consider   path integration  of a fermionic oscillator with a
 one-parameter family of  boundary  conditions with respect to the 
time coordinate. 
The dependence of the fermion determinant on these boundary conditions is 
derived in a closed form  with the help of the self-adjoint extension of
differential operators. 
The result reveals  its crucial dependence on them, contrary to the conventional 
understanding that  this  dependence becomes  negligible over  sufficiently
long time  evolution.
An example in which   such dependence plays a significant role is discussed 
in  a model of supersymmetric quantum mechanics.
\end{abstract}

PACS: 11.30.Fs;  11.30.Pb; 03.65.-w; 02.70.Hm;

Keywords:  path-integral; boundary condition;  supersymmetric quantum mechanics; 
\vspace{0.5cm}

\newcommand{\D}{{\cal D}}
\newcommand{\be}{\begin{equation}}
\newcommand{\ee}{\end{equation}}

The determinant arising  from fermionic path integral over Grassmann
variables brings us important and critical  information on  models  in  field theory.
Anomalies \cite{Adler, Adler.Bardeen, Witten1, Alvarez} in gauge theories are 
a typical  example of  this kind.
They reveal   that the symmetry  of  the classical
action breaks down when the  fermion fields are  integrated out to yield the
determinant \cite{Fujikawa1}.
They critically restrict the model in its fermion content  \cite{Alvarez}.
Thus any question about  the determinant deserves  careful investigation. 
The issue of  this letter is the dependence of the determinant
on  boundary conditions with respect to the  time coordinate.

It is useful to formulate fermions in field theory by putting  the system in a 
finite spatial box.
Owing to this treatment the one-particle states can  be labeled with an integer.
To each one-particle state,  we assign a fermionic oscillator  to
distinguish   the  vacant  and   occupied states.
The fermions are then described by a set of the oscillators which interact mutually  
through their coupling  to background bosonic degrees of freedom. 
For example, in Euclidean path integral formalism,  a chiral two-component  fermion field in a gauge theory  is defined  by the Lagrangian 
\begin{equation}
 L = \sum_{a,b} \left[ \bar\psi_a {d \over d\tau} \psi_a +  \bar\psi_a  \left[ \nu_{ab}(\tau)  + i \omega_{ab}(\tau) \right] \psi_b\right], \label{1}
\end{equation}
where \(\bar\psi_a\) and \(\psi_a\) are anti-commuting fermionic variables
 which  correspond to the creation and annihilation operators, respectively,  of  the
\(a\)-th one-particle state,    \(\nu_{ab}\)  is the matrix element of  the Hamiltonian 
\( i [ \vec \partial +i \vec  A(x) ] \vec \sigma\)  between the \(a\)-th and the \(b\)-th one-particle states.   Similarly,  \(\omega_{ab}\)  arises from the temporal component of 
the gauge field.   
To fully comprehend fermions in field theory,
we will explicitly calculate the determinant of a  fermionic oscillator, i.e.
a fermion with only one one-particle state,    with a variety of
boundary conditions.

In the following, we restrict the  frequency of the oscillator,  \(\nu(\tau) + i \omega(\tau) \),
which is the one-component counterpart of the matrices in Eq.~(\ref{1})  
to be real. This imitates a calculation in the temporal gauge (\(\omega = 0\)).
Since the frequency is real,   two quantum states, vacant and occupied, of the oscillator 
have  real and positive-definite transition amplitudes for  imaginary time evolution.
The path integral over  anti-periodic configurations is
well known to yield the sum of the amplitudes of the two states.  
Thus the determinant is real and positive-definite.
The periodic boundary condition also yields the sum of their amplitudes,  but in which 
each one  is weighted by \( (-)^F\),  where \(F\) denotes the occupation number 
of the corresponding states  \cite{Cecotti}. 
If the amplitude of the occupied state is larger than that of the vacant, 
the determinant becomes negative. 
In a  previous study \cite{Kikuchi1},  we  considered   the determinant 
with  a one-parameter family of  boundary conditions 
which includes the periodic and  anti-periodic conditions.
There, our motivation  was  to clarify the role of a  zero-mode in  the path integral.  
We have shown the boundary condition which  admits it  is not  the same as 
the anti-periodic condition and is  generally different  from the periodic one. 
The determinant becomes zero under  this condition.

Naively,  one might think that the  dependence of the determinant on boundary conditions
becomes negligibly  small  as  the  time interval goes to infinity.
The difference in the values of the determinant,  however,  does not disappear.
We extend  the calculation to a wider family  of boundary conditions than we did in
the previous study \cite{Kikuchi1}
by letting the parameter be complex,  thus confirming its dependence.
We will also provide an example in supersymmetric quantum mechanics that shows
a crucial dependence on the boundary conditions when analyzed in path integration.

The path integral of the fermionic oscillator  is  given by
\be
I  =  \int [d\bar\psi d\psi ] \exp
\left[-\int_{0}^{T} d\tau \bar\psi \D \psi \right],\label{I}
\ee
where \( {\cal D} = d/d\tau + \nu(\tau)\) and \(\nu(\tau)\) is the time-dependent {\em real}  
frequency  induced by background bosonic degrees of freedom.
The  integration over the anti-commuting variables, \(\psi\) and \(\bar\psi\),
is implemented  by  Berezin 
 \cite{Berezin,Faddeev, Onuki, Coleman},  and the result is called the
determinant of \(\D\).
Since  \(\D\) is  {\em not}  a finite-dimensional matrix, its determinant is  {\em not}
simply the  product of its  eigenvalues.
Difficulty arises because \(\D\) is not hermitian: 
we use the inner product of  vectors, say \(\varphi\) and \(\phi\), defined by 
\(
(\phi, \varphi) \equiv \int_{0}^{T}d\tau \phi^* \varphi
\) 
in the Hilbert space composed of square-integrable functions of \(\tau\) in the
interval \([0, T]\);  the adjoint  \(\D^\dagger = -d/d\tau + \nu(\tau) \)
is then different from \(\D\).
We do not know a proper setup in the eigenvalue problem of 
non-hermitian operators 
that provides us with such a useful tool for path integral as a complete orthonormal set of
eigenvectors.
Even the meaning of  the eigenvalues of \(\D\)  is not very clear:
under the boundary conditions we deal with in the following,
the  linear manifold  on which \(\D\) operates  is generally different  from
that which  results  from  its operation.

Fujikawa has proposed a  method  which is useful  in this situation \cite{Fujikawa2}.
Following him,  we solve the eigenvalue problems of \(\D^\dagger \D\) and \(\D \D^\dagger\).
Appropriate boundary conditions make these operators self-adjoint and non-negative.
Then their normalized eigenvectors
 \(\varphi^{(n)} \) and \(\phi^{(n)} \) (\(n = 1, 2, 3, ....) \),
\be 
\D^\dagger \D \varphi^{(n)} = \lambda_n \varphi^{(n)},\quad 
\D \D^\dagger \phi^{(n)} = \lambda_n \phi^{(n)},\label{defdiff}
\ee
have a one-to-one correspondence to each other, 
\be {1 \over \sqrt \lambda_n } \D \varphi^{(n)} = \phi^{(n)}, \quad
{1 \over \sqrt \lambda_n } \D^\dagger \phi^{(n)} = \varphi^{(n)}.
\label{1T1}\ee
This relation  holds  for all pairs of eigenvectors with a positive eigenvalue,  and
their spectra, the sets of eigenvalues,  are identical except   a possible difference at
zero eigenvalue.
Using the expansions by these eigenvectors,
\be
\psi(\tau) = \sum_n a_n \varphi^{(n)}(\tau), \quad 
\bar\psi(\tau) = \sum_n \bar a_n \phi^{(n)*}(\tau) ,\label{anan}\ee
we transform the integral measure  \([ d\bar \psi\, d\psi]\)  to \([d\bar a\,d a]\) 
and  obtain
\be 
I = {\cal N} \int  \prod_n [d\bar a_n d a_n ] 
e^{ - \sum_n \sqrt \lambda_n \bar a_n a_n }
 = {\cal N} \left[ \det(\D^\dagger \D) \right]^{1/2}, \label{Irst}
\ee
where  \(\cal N\) is the Jacobian between the two measures,  and
\([\det(\D^\dagger \D)] ^{1/2} \equiv \prod_n \sqrt \lambda_n\) 
is the infinite product of the square-root of the eigenvalues.
Although the latter is divergent,  it gives a finite result  to \(I\)  
when combined with \(\cal N\).

The boundary conditions we use are parametrized  by  one complex variable \(\beta\).
The conditions  for \(\D^\dagger \D\) are  given by
\be  \varphi(0) + \beta \varphi(T) = 0, \quad \beta^* \D\varphi(0) + \D\varphi(T)
=0, \label{Bvarphi}\ee
and those for  \(\D\D^\dagger\) are  given by
\be  \beta^* \phi(0) + \phi(T) = 0, \quad \D^\dagger \phi(0) + \beta \D^\dagger \phi(T)=0.\label{Bphi}\ee
Note that these conditions keep the one-to-one relation between
\(\varphi\) and \(\phi\) in Eq.~(\ref{1T1}).
They define  two linear sub-manifolds in the Hilbert  space.
These linear manifolds  are  the  domains   of  \(\D^\dagger \D\) and \(\D \D^\dagger\),
respectively%
\footnote{ We refer to a linear sub-manifold on which a linear  operator is defined to act as
a domain.}.
We can readily verify  equations
\( (\D^\dagger \D \varphi_1, \varphi_2 ) = (\D \varphi_1, \D \varphi_2) = 
(\varphi_1, \D^\dagger \D \varphi_2) \)
for  \(\varphi_1\) and \(\varphi_2\) which are arbitrarily chosen in the domain of
\(\D^\dagger \D\).
Thus \(\D^\dagger \D\) is at least  symmetric and non-negative;
and so is \(\D\D^\dagger\). 
They are  in fact   self-adjoint  so as to have a complete  orthonormal basis of eigenvectors,  and  the expansions in  Eq.~(\ref{anan}) cover all possible configurations. 
The proof is given based on  the mathematical theory  of the self-adjoint  extension of 
 symmetric operators \cite{Akhiezer};   it  will be detailed  in a separate publication  \cite{Kikuchi2}.

A different value of \(\beta\) provides a different domain for \(\D^\dagger \D\).
Thus  its determinant depends on \(\beta\)  as well as the
background \(\nu(\tau)\),  \(I = I_\beta[\nu] \).

We calculate the modulus of \(I\) first and its phase next.
However, prior to  that, we  view  how the determinant has 
 phase, since  it is  helpful to have a perspective on our  calculations.
Let us start  at Eq.~(\ref{1T1}). 
We always take the positive root of \(\sqrt \lambda_n\) in this equation. 
The relative phase between \(\varphi\) and \(\phi\) is then fixed.
The consequent determinant  \([\det( \D^\dagger \D)]^{1/2}\) in Eq.~(\ref{Irst})
becomes a product of positive quantities,  and  it does not induce  phase.
Hence only  the Jacobian \(\cal N\) contains  the phase of \(I\).

Now assume  temporarily that \(\beta\) is real.  
We can then choose   real functions  for the eigenvectors \(\varphi^{(n)}(\tau) \) 
and \(\phi^{(n)}(\tau)\).
These functions can be regarded as the ``transformation matrix'' between 
\(\psi(\tau) (\bar\psi(\tau))\) and  \(a_n (\bar a_n)\),  according to  Eq.~(\ref{anan}).
The Jacobian \(\cal N\) is  related to their  ``determinant" and  is  real for this case.
At  \(\beta = 1\)  the determinant is positive-definite, and 
thus the phase in \(\cal N\) is zero independently of  \(\nu(\tau)\).

For a complex value of \(\beta\),  the eigenvectors cannot be  real
and \(\cal N\) becomes complex. 
To obtain its phase, 
we compare two different measures,  one  is related  to the coefficients in expansion  
(\ref{anan})  where the  eigenvectors are solved for the complex \(\beta\),  and the other 
is obtained from those for \(\beta = 1\)  but with the same frequency  \(\nu(\tau)\).
The value of \(\cal N\)  differs depending on to which measure  
the original one  \([d\bar\psi d\psi]\) is transformed.
Let \(\varphi^{(n)}{}'\) and \(\phi^{(n)}{}'\) be  normalized eigenvectors 
 and \([d\bar b d b]\) be their related measure for  the value of \(\beta\) of interest;
similarly  \(\varphi^{(n)}\),  \(\phi^{(n)}\) and \([d\bar a d a]\) are obtained  for  \(\beta = 1\).
These  measures  are related to each other  
by  \( [d\bar a da ] = {\cal J} [d \bar b d b] \),  where
\be
{\cal J} \equiv [ \det(\varphi^{(n)}, \varphi^{(m)'} )]^{-1}\times [\det (\phi^{(n)*}, \phi^{(m)' *})]^{-1}.
\label{J}
\ee
The expressions \((\varphi^{(n)}, \varphi^{(m)'} )\)  and  \( (\phi^{(n) }{}^*, \varphi^{(m)'*})\)
in Eq.~(\ref{J}) are  the inner product of the \(n\)-th and the \(m\)-th eigenvectors;
each of them is  the element of the transformation matrix  
between the two different orthonormal complete bases. 
The power \((-1)\) in Eq.~(\ref{J})  comes from the fact that the measure is for 
the integration of  Grassmann variables \cite{Berezin}.
Since the bases are complete and orthonormal, the transformation matrices between them
 are unitary, and \(\cal J\) is  a phase factor.
\(\cal N\) changes by \(\cal J\) when the different bases are used, 
and thus the  phase in \(\cal N\) is that of \(\cal J\),
which is also the phase of \(I\).

Note that each determinant factor in Eq.~(\ref{J})  is in fact not well-defined separately.
The eigenvectors  have arbitrariness in their phase,
but  each factor  is not invariant under the change  in the phase.
It is the  one-to-one correspondence between \(\varphi\) and \(\phi\), Eq.~(\ref{1T1}),  that 
correlates the phase of \(\varphi\) to that of \(\phi\)  and makes \(\cal J\) 
invariant despite this arbitrariness.

\newcommand{\ampvac}{{\cal M}_0}
\newcommand{\ampocc}{{\cal M}_1}

The modulus of the determinant  is readily obtained  by slightly extending the calculation 
 in the previous study  \cite{Kikuchi1} to the case where  the  parameter \(\beta\) can be 
complex.
To obtain the modulus,  we define a two-by-two matrix
\be M_\beta (z) \equiv \left(\begin{array}{cc}
u_1(z; 0) + \beta u_1(z; T) & u_2(z; 0) + \beta u_2(z; T) \\
\beta^*  \D u_1(z; 0) + \D u_1(z; T) & \beta^* \D u_2(z; 0) + \D u_2(z; T)
\end{array}
\right),\label{Mmp}\ee
where \(u_1\) and \( u_2\) are
two linearly independent solutions of   a \(z\)-parametrized differential equation
\be 
\D^\dagger \D u_i(z; \tau) = z u_i(z;\tau),  \label{Eqforu}
\ee
solved with initial conditions
\( u_1(z; 0) = 1 \),  \( \dot u_1(z;0) \equiv d u_1(z; 0)/ d\tau = 0\), 
\( u_2(z;0) = 0 \), and \( \dot  u_2(z;0) = 1\).
The complex parameter \(z\) in Eq.~(\ref{Eqforu}) becomes  an eigenvalue of 
\(\D^\dagger \D\) if and only if  some linear combination of \(u_1\) and \(u_2\) 
 meets the boundary conditions in  Eq.~(\ref{Bvarphi}).
In other words, there exists  a two-component non-zero vector \(\gamma_i\) that satisfies
 \([M_\beta (z)]_ {ij} \gamma_j = 0\).
This is why we choose the particular form  for \(M_\beta\) in Eq.~(\ref{Mmp}).
The condition  that  the \(z\) is  an eigenvalue  of \(\D^\dagger \D\) is equivalent to
\(\det M_\beta (z) = 0 \).  

From this behavior of \(M_\beta\),  we can prove the identity 
\be
{\det (\D^\dagger \D -z) \over \det (\tilde\D^\dagger\tilde \D -z)} 
\equiv \prod _{n = 1}^\infty \left( {\lambda_{n} -z \over \tilde \lambda_{n}- z} \right)
 = {\det M_\beta(z) \over \det \tilde M_\beta (z)}
\label{iddet}
\ee 
where the tildes  used here  distinguish two different operators,  matrices and eigenvalues 
which are defined 
with different frequencies, say \(\nu(\tau) \) and \(\tilde \nu(\tau)\),  
but with the same value of the  parameter \(\beta\).
Two fractional expressions, far left and far right  in (\ref{iddet}),  are  meromorphic functions of \(z\). 
Owing to the behavior of \(M_\beta\), 
they  have  poles and zeros at  same values of \(z\), i.e.  
zeros at eigenvalues of \(\D^\dagger \D\)
and poles at those of \(\tilde\D^\dagger \tilde \D\).
Both  functions converge to \(1\) asymptotically   as  \(|z|\)  goes to infinity except along
the real positive axis \cite{Kikuchi1}.
The ratio of the two meromorphic functions is thus an analytic  function of \(z\)
that goes to \(1\) as \(z\) goes to infinity in any direction, and is necessarily a constant \(1\)%
\footnote{The same argument is used in Ref.~\cite{Coleman} to calculate a different determinant.}.
This concludes  the proof of  Eq.~(\ref{iddet}).

The identity (\ref{iddet})  means that  the ratio  \( | {\cal N} [\det(\D^\dagger \D)]^{1/2} | / [\det M_\beta(0)]^{1/2} \) does not depend on the \(\nu(\tau)\) which is used to define
 \(\D\) and  \(M_\beta\).
We can thereby obtain  the functional dependence of the modulus on \(\nu(\tau)\)
by calculating  the determinant of \(M_\beta\). 
Putting  the solutions at \(z = 0\), 
\begin{eqnarray}
u_1(0;\tau) &=&  \exp\left[ - \int_{0}^\tau d\tau' \nu(\tau') \right]\times
\left\{1 + \nu(0) 
\int_{0}^\tau d\tau' \exp\left[  2 \int_{0}^{\tau' }d\tau'' \nu(\tau'') \right] \right\},\nonumber \\ 
u_2(0;\tau) &= &  \exp\left[ - \int_{0}^\tau d\tau' \nu(\tau') \right] \int_{0}^\tau d\tau' 
\exp\left[  2 \int_{0}^{\tau' }d\tau'' \nu(\tau'') \right],
\label{us}
\end{eqnarray}
into the expression of \( M_\beta(0)\) in Eq.~(\ref{Mmp}) and taking its determinant,  we obtain
\be
| I | = {\cal N}' \left[  (\ampvac + \beta \ampocc)(\ampvac + \beta^* \ampocc) \right]^{1/2},
\label{Imodu}
\ee
where \(\ampvac =    \exp\left[ (1/2)  \int_{0}^{T} d\tau \nu(\tau) \right]\),
\(\ampocc =    \exp\left[ -(1/2)  \int_{0}^{T} d\tau \nu(\tau) \right]\) and
\(\cal N'\) is a  positive constant  that may depend only on \(\beta\).

Eq.~(\ref{Imodu}) implies \(I\) vanishes  for \(\beta = - {\cal M}_0{}^2\)  
 (note  \({\cal M}_1 = {\cal M}_0^{-1}\)).
This is because the boundary conditions with this value of \(\beta\) admit a zero mode
in the domain.
Its effect on the determinant calculation has been clarified  in \cite{Kikuchi1}.

We next proceed to the calculation of the phase,  \(\ln \cal J\).
Our plan  is to integrate infinitesimal variations induced  in  the phase
while  the parameter \(\beta\) moves continuously from \(1\) to some value of interest.
 \(\nu(\tau)\)  is kept fixed in this process.
We use the  formula \(\delta \ln \det M = {\rm Tr} M^{-1} \delta M \) that holds for an infinitesimal
variation of any matrix \(M\).
Applying this formula  to the variation  \(\delta \ln {\cal J}\) with \(\cal J\) in  Eq.~(\ref{J}), 
and using the fact that  a  set of eigenvectors is  complete,
we obtain
\be 
\delta \ln {\cal J} = \sum_n \left[ (\phi^{(n)},  \phi^{(n)'}-\phi^{(n)}) 
-  (\varphi^{(n)}, \varphi^{(n)' } - \varphi^{(n)}) \right]  = \sum_n \left[ (\phi^{(n)},  \phi^{(n)'}) 
-  (\varphi^{(n)},  \varphi^{(n)'}) \right], \label{dlJ}
\ee
where \(\varphi^{(n)}\) and  \(\phi^{(n)}\) are eigenvectors  at \(\beta\) and
primed ones are those at \(\beta + \delta \beta\).
We put  the one-to-one correspondence \(\phi^{(n)} = \D\varphi^{(n)}/\sqrt{\lambda_n}\) 
into (\ref{dlJ}).  After  integrating by parts  and using the similar one-to-one relation for \(\phi^{(n)'}\) 
and the boundary condition (\ref{Bvarphi}), we obtain
\be
\delta \ln {\cal J} = \sum_n {1\over \lambda_n} \left[ {\delta \beta^* \over \beta^*} \varphi^{(n)*}({T}) \D \varphi^{(n)}({T}) + {1\over 2} \delta \lambda_n
 \right],
\ee
where \(\delta \lambda_n \) is the variation of the \(n\)-th eigenvalue under the infinitesimal change of \(\beta\). 
\(\delta \lambda_n \) is calculated by the relation \(\delta\lambda_n (\varphi^{(n)}, \varphi^{(n)'}) = 
(\varphi^{(n)}, \D^\dagger \D \varphi^{(n)'}) - ( \D^\dagger \D \varphi^{(n)}, \varphi^{(n)'})\) as
\be 
\delta\lambda_n = - {\delta\beta^*\over \beta^*} \varphi^{(n)*}({T})\D\varphi^{(n)}({T}) -  {\delta\beta\over \beta} \D\varphi^{(n)*}({T}) \varphi^{(n)}({T}).
\ee 
We then  obtain
\be
\delta \ln {\cal J} = \sum_n {1\over 2 \lambda_n} \left[ {\delta \beta^* \over \beta^*} \varphi^{(n)*}(T) \D \varphi^{(n)}({T})  -  {\delta\beta\over \beta} \D\varphi^{(n) *}(T) \varphi^{(n)}(T)
 \right].  \label{lnJ2}
\ee
This  expression manifestly shows  that  \( \delta \ln {\cal J} \) is  purely  imaginary  as it should be.

To sum up the terms in Eq.~(\ref{lnJ2}),  we use 
the resolvent \( R_z \equiv (\D^\dagger \D -z )^{-1}\),  which has the expression
\be 
R_z(\tau,\sigma) = \sum_n{1\over \lambda_n -z} \varphi^{(n)}(\tau) \varphi^{(n)*}(\sigma)
\ee
as an integral kernel.
We notice readily that  contour integration of \( \D R_z  / z \) over \(z\) along a contour  that  goes around all  eigenvalues of  \(\D^\dagger \D\) clockwise in the complex \(z\)-plain gives 
the relevant part of the sum.
The contour we use here is  one that 
goes  below the real positive axis towards the origin from  infinity until 
it passes all the eigenvalues  and then goes back moving  above the real axis. 
Without changing the value of the integral,  we can add to it
another contour  that almost makes  a circle at infinity 
but does not  cross  the real axis,  because the value of the  integral along the latter contour is zero.
We now have the  integral along  the closed contour \(C\) (see Figure 1), 
in which only  the pole at the origin  contributes to the integral.
We obtain
\be
\delta \ln {\cal J} = {1\over 2}\left[ \left. {\delta \beta^*\over \beta^*} \D R_0(\tau, \sigma)\right|_{\tau = \sigma = T}
 - ({\rm c.c.}) \right]
\label{dlj2}
\ee
as the result of the sum.

We have  assumed that \(\beta\) always stays off the value 
at which one eigenvalue becomes zero  and \(R_0\)  becomes singular; 
the phase of \(I\) is  obviously meaningless  when \(I\)   is zero.

In order to calculate the resolvent \(R_0\) in Eq.~(\ref{dlj2}),
we have to know some mathematical details  about the self-adjoint extension of
differential operators.
Following  a  textbook \cite{Akhiezer},
we  briefly describe a specific  extension  of  the operator \(\D^\dagger \D\) 
 to the extent where our calculation of  \(R_0\) appears convincing;  
the self-contained description will be given in \cite{Kikuchi2}.
The operator that is  extended to be self-adjoint is \(\D^\dagger \D\)  defined, however,   in 
a domain \( D_0\)  under  more restrictive boundary conditions than Eq.~(\ref{Bvarphi}).
These conditions  are \(\varphi(0) = \varphi(T) = \dot\varphi(0) =  \dot\varphi(T) = 0\).
\(\D^\dagger \D\) defined in \(D_0\) is  symmetric but is not self-adjoint:  
the domain of its adjoint  is not restricted by a boundary condition and is obviously larger than 
 \( D_0\).
We have to extend \( D_0\) by adding  {\em two} appropriate vectors  in order to
make  \(\D^\dagger \D\) self-adjoint, that is,  the operator itself  and its adjoint have the same domain.
These two vectors have the form
\be 
w_i(z) = U_{ij} u_j(z^*) - u_i(z)\quad  ( i, j = 1, 2 ), \label{w}
\ee
where  \(u_i\) is the solution to  Eq.~(\ref{Eqforu})  with  an arbitrary  but  
non-real parameter \(z\),  and  \(u_i(z^*) = u_i^*(z) \) is  its complex conjugate.
The two-by-two matrix \(U\) in Eq.~(\ref{w})  has to be ``unitary"  in the sense that
the mapping defined by  \(u_i \rightarrow U_{ij} u^*_j \)
from the two-dimensional  vector space composed of \(  u_i \)s  to that of  \(u^*_i\)s  is unitary.
\(\D^\dagger \D\) becomes self-adjoint  in this extended domain \(  D \equiv  D_0  \oplus \{ w_1, w_2\} \)  \cite{Akhiezer}.

The  unitarity of \(U\) solely does not  determine  \(U\),  but boundary conditions do.
The  requirement that any vector in \(D\) should meet  the  boundary conditions  in Eq.~(\ref{Bvarphi})
solves  \(U \)  as 
\be 
[U_\beta(z) ]_{ij}  =  \left[ M_\beta(z^*)\right]^{-1}_{jk} \left[ M_\beta(z)\right]_{ki}
\label{Ubz}\ee
for the  parameters  \(\beta\) and \(z\). 
The  mapping  induced by this particular \(U_\beta(z) \) is  shown to be unitary \cite{Kikuchi2}.
The self-adjoint operator \(\D^\dagger \D\) that we have  used is in fact this extended 
operator.

Having described  the structure of the domain \(D\),
we can now resume the calculation of  the resolvent.
What we  need to know in Eq.~(\ref{dlj2}) is its value only at the end of the time interval.
Among  all the vectors in \(D\),  only the two  \(w_i\)s in Eq.~(\ref{w})  can contribute  to
\(R_z\) at the end.
We examine the operation of  \((\D^\dagger \D - z) \) on \(w_i\) and find
\be
 (\D^\dagger \D -z) w_i(z)  = (z^* -z ) [U_\beta(z)]_{ij} u_j(z^*), \label{DDw}
\ee
which enables us to obtain the necessary information on \(R_z\).
Since zero is not an eigenvalue,  we can take  the  limit \(z \rightarrow 0\) in  Eq.~(\ref{DDw})
to obtain   \(R_0  u_i = \tilde w_i\),  where
\be
\tilde w_i \equiv \lim_{z \rightarrow 0} {w_i(z) \over z^* - z} =
 \lim_{z \rightarrow 0} {1 \over z^* - z}\left\{  [U_\beta(z)]_{ij} u_j(z^*) - u_i(z)\right\},
\label{wtilde}
\ee 
and we have used \(U_\beta(0) = 1\) (see Eq.~(\ref{Ubz})). 
As an integral kernel, \(R_0\)  is written as 
\be
R_0(\tau,\sigma ) = \sum_{i = 1,2} \tilde w_i(\tau) \tilde  u_i(\sigma) + ... \label{R0},
\ee
where \(\tilde u_i\)s are   the  linear combinations  of \(u_i\)s  that  satisfy 
 \( (\tilde u_i, u_j) = \delta_{ij}\), and we have excluded  terms that  do not contribute to the final result.
Both \(\tilde w_i\) and \(\tilde u_i\) can be explicitly obtained  with the solutions \(u_i\)s  at \(z = 0\).
Plugging these results into (\ref{dlj2}) and after a long but straightforward calculation, 
we  obtain
\be
\delta \ln {\cal J} = {1\over 2}  \left[ { {\cal M}_1  \delta \beta \over {\cal M}_0 + \beta {\cal M}_1 }
-  {{\cal M}_1  \delta \beta^* \over {\cal M}_0 + \beta ^* {\cal M}_1 }
 \right].
\ee
By  integrating  this equation under the condition   \({\cal J} = 1\) at \(\beta = 1\),  we finally obtain 
\be
{\cal J} = \left[  { {\cal M}_0 + \beta {\cal M}_1 \over {\cal M}_0 + \beta^* {\cal M}_1 }\right]^{1/2}.
\label{Jrst}
\ee

The dependence  of the phase factor  in Eq.~(\ref{Jrst})  on the parameter \(\beta\) 
shows the advantage of the present calculation where we have allowed
\(\beta\)  to  be complex.  
Let \(\beta\) move along the real axis, for example,  from \(1\) to \(-1\) and assume 
the value \(  - {\cal M}_0{}^2 \) is between them.
At  this  point  the determinant becomes zero,  and there will occur
an ambiguity in  its sign after \(\beta\) passing the point  if we know only its
modulus.
This ambiguity is solved since we can use a path   
that circumvents the point  along an circle with infinitesimal radius \(\epsilon\) as 
\( \beta = - {\cal M}_0^2 + \epsilon e^{i \theta} \)  where \(\theta\) moves from \(0\) to \( \pi \).
The determinant becomes negative after \(\beta \)  passes the value.

Combining the modulus (\ref{Imodu}) and the phase factor (\ref{Jrst}), we obtain 
\be
I_\beta[\nu] = {\cal N'} \left( {\cal M}_0 + \beta {\cal M}_1 \right) \label{IrstF}.
\ee
\({\cal M}_0\) and \({\cal M}_1\) are in fact the transition amplitudes for 
the vacant and the occupied states, respectively.
The result in Eq.~(\ref{IrstF}) is  not only consistent with  the known case of  
the anti-periodic (\(\beta = 1\)) and the periodic (\(\beta = -1\))  boundary
conditions, but also applies to  any complex value of \(\beta\).
\(\cal N'\) is an overall normalization and   will not affect the following discussion.

Although we have considered only  a single fermionic oscillator,  our result  indicates that  
 the similar  dependence is expected to appear 
in  fermion determinants in field theory,  perhaps in a more complicated  manner.
At least for a simple case in which  there are a finite number  of fermionic oscillators  that share 
 a common value of  \(\beta\) for their boundary conditions,  
we can safely conclude, without seeing anomalies, 
that the determinant is the trace of the time evolution operator in which the amplitude of each state is weighted by \(\beta^F\),  
according  to its fermion number \(F\).
It is hard to imagine that the difference of  the determinant  caused by  
such  a non-trivial weighting  disappears even in the limit \(T \rightarrow \infty\).
Thus we have to be careful which  boundary condition we use in the calculations.

In order to  see how the boundary conditions affect  the result,  let us examine  an  illuminating example
--- the supersymmetric double-well quantum mechanics \cite{Witten2}.
With  \(q\) and \(p\),  the coordinate and momentum of   bosonic degrees of freedom,
  \(\psi^\dagger \) and \(\psi \),   the creation and annihilation operators 
for  fermion, we write the  Hamiltonian of this model as
\be
H = {1\over 2} \{ Q, Q^\dagger\} = {1\over 2} (p^2 + W(q)^2) +{1\over 2} {d W(q)\over dq} 
[\psi^\dagger, \psi],
\ee
where \(Q\equiv ( p + iW(q)) \psi\),   \( Q^\dagger \equiv ( p - i W(q)) \psi^\dagger \)
 and  \(W\) is chosen to be
\(W(q) = q ( 1- g q) \) with a coupling constant \(g\).
What will  happen if one analyzes the model by  path integral without paying
 attention to  the boundary conditions?
One  may try to integrate the fermion first in a given bosonic background 
and then to  integrate  the bosonic part  weighting the bosonic measure with
 the fermion determinant,  which  is given by Eq.~(\ref{I})  with \(\nu(\tau) =  1 - 2 g q(\tau)\).
Under  the periodic boundary condition,   the determinant 
changes the sign when  the background varies between \(q(\tau)\) and \(1/g - q(\tau)\), 
for which  the bosonic potential \(W(q)^2\)  takes the same  value. 
The contribution of the two configurations  cancels each other out, and the path integral becomes  zero. 
One may  conclude that the model is ill-defined  by interpreting  this result 
as the statistical trace being zero.

We  know, however,  that the model is well-defined, and further it  is an  instructive 
example of the dynamical supersymmetry breaking \cite{Witten2}. 
The path integral over  the periodic configurations  is not the statistical  trace,  
but the regularized Witten index \({\rm Tr} (-)^F e^{-TH} \) \cite{Witten3, Cecotti}.
The fact that it is zero  merely implies  the possibility that two perturbative  zero-energy states, 
one localized at \(q = 0\) with  \( F = 0 \) \((F\equiv \psi^\dagger \psi)\) and the other 
localized at \( q = 1/g\)  with \(F = 1\),
may be lifted in pairs  by non-perturbative effect of the interaction, 
and the supersymmetry may be broken.
It indeed happens in this model by the valley-instanton effect \cite{Aoyama.et.al}.

This example shows that in any model which exhibits dynamical supersymmetry
breaking,  its  fermion determinant needs to have non-trivial dependence on
the boundary conditions.
The path integral  has to vanish in the case of the periodic fermion fields  so that the Witten index is zero,
while it  should provide a non-vanishing value  for anti-periodic fermion fields in order that
the partition function is not zero.
Thus the careful study of the dependence of fermion determinants on the boundary conditions
is   important in the search for a model in which supersymmetry breaks dynamically.

\begin{center}
{\bf Acknowledgments}
\end{center}
The author  thanks  H.~Aoyama  for the  discussion  on the subject  and his 
comments on the manuscript,  K. Funakubo and M.~ Sato  for  the  discussions
with them.
He also thanks M.~Itoh for the critical reading of the manuscript and
his useful suggestions about  English usage. 
D.~Jones and T.~Fukube  gave  the author  helpful comments on the English in the manuscript,
which he appreciates as well.

\newcommand{\J}[4]{{\sl #1} {\bf #2} (19#3) #4}
\newcommand{\MPL}{Mod.~Phys.~Lett.}
\newcommand{\NP}{Nucl.~Phys.}
\newcommand{\PL}{Phys.~Lett.}
\newcommand{\PR}{Phys.~Rev.}
\newcommand{\PRL}{Phys.~Rev.~Lett.}
\newcommand{\AP}{Ann.~Phys.}
\newcommand{\CMP}{Commun.~Math.~Phys.}
\newcommand{\CQG}{Class.~Quant.~Grav.}
\newcommand{\PRP}{Phys.~Rept.}
\newcommand{\SPU}{Sov.~Phys.~Usp.}
\newcommand{\RMPA}{Rev.~Math.~Pur.~et~Appl.}
\newcommand{\SPJ}{Sov.~Phys.~JETP}
\newcommand{\MP}{Int.~Mod.~Phys.}
\newcommand{\JMP}{J.~Math.~Phys.}
\newcommand{\PTP}{Prog.~Theor.~Phys.}
\newcommand{\NC}{Nuovo Cim.}

\newpage
Figure caption:
Figure 1.   The schematic drawing of  the contour \(C\) in the complex \(z\)-plane of \( \D R_z  / z \).
\(C\) is a closed contour made of a contour that goes around all eigenvalues of \(\D^\dagger \D\)
clockwise and another that goes around at infinity counterclockwise, as explained in the text.
The crosses indicate  positions of the poles of \( \D R_z  / z \).
\end{document}